# Load Shaping Based Privacy Protection in Smart Grids: An Overview

*Cihan Emre Kement* [1, 2]

### Abstract

*Fine-grained energy usage data collected by Smart Meters (SM) is one of the key components of the smart grid (SG). While collection of this data enhances efficiency and flexibility of SG, it also poses a serious threat to the privacy of consumers. Through techniques such as nonintrusive appliance load monitoring (NALM), this data can be used to identify the appliances being used, and hence disclose the private life of the consumer. Various methods have been proposed in the literature to preserve the consumer privacy. This paper focuses on load shaping (LS) methods, which alters the consumption data by means of household amenities in order to ensure privacy. An overview of the privacy protection techniques, as well as heuristics of the LS methods, privacy measures, and household amenities used for privacy protection are presented in order to thoroughly analyze the effectiveness and applicability of these methods to smart grid systems. Finally, possible research directions related to privacy protection in smart grids are discussed.*

## 1. INTRODUCTION

Rapid changes in the technology towards the end of the 20th and the beginning of the 21st century imposed changes in the people's lives as well as their energy consumption. Daily life becomes more and more dependent on electricity, due to the digital transformation and new products such as plug-in electric vehicles (PEVs). Today's electric grid is expected to meet the increasing consumer demand as well as to be more reliable, efficient, scalable and secure. In order to fulfill these expectations, *smart grids* (SGs) are being developed, where the energy flow is coupled with a two-way information flow between the components of the grid structure.

The information flow throughout the SG has many advantages. Distribution system operators (DSOs) can quickly detect failures or shortages, whereas electricity generators (EGs) can monitor the electricity usage data in order to tune electricity production and hence increase efficiency. The latter is made possible by *advanced metering infrastructure* (AMI), where fine-grained electricity usage of consumers is collected with the help of smart meters (SMs).

Although the high-granularity consumption data is very useful for both the supply and demand sides, it also contains very detailed information about the consumers' private lives. By using nonintrusive appliance load monitoring (NALM) techniques, the power signature of a house can be disintegrated into the individual appliances [1]-[7]. Therefore, sensitive information such as the number of residents, daily consumption patterns of the residents and whether the house is vacant or not can be inferred from the collected data.

In order to protect the consumers' privacy, many methods such as load shaping (LS), adding noise to the consumption data, differential privacy, data aggregation and cryptographic measures have been proposed in the literature. Among these, LS methods stand out as they hide the details of electricity usage while providing the utility company (UC) with the true consumption data.

The remainder of this paper is as follows: First, some background information about the AMI and NALM techniques, as well as different notions of privacy in the case of SGs are provided. Then, various privacy protection methods are categorized, indicating their advantages and disadvantages, followed by the LS methods in detail. The means used for LS are explained, along with their limitations. Then, various privacy measures proposed are reviewed. Finally, future research directions for LS based privacy protection in SGs are proposed. Concluding remarks are provided in the conclusion section.

[1] *Laboratory of Information and Decision Systems (LIDS), Massachusetts Institute of Technology, Cambridge, MA, 02139, USA.* kement [at] mit.edu

[2] *TOBB University of Economics and Technology, Department of Electrical and Electronics Engineering, 06510, Cankaya, Ankara, Turkey.* ckement [at] etu.edu.tr

## 2. BACKGROUND INFORMATION

### 2.1. Advanced Metering Infrastructure (AMI)

AMI is an integrated system of communication networks and SMs which enables the two-way communication between consumers and UCs [8]. This two-way communication comes with many advantages. It enables DSOs to automatically detect outages, which would be otherwise reported by customers. UCs can learn a lot more detail about the electricity usage trends of their consumers. By using this information, UCs can increase their efficiency. Furthermore, UCs can provide real-time electricity prices to consumers, who can use this information to reduce their electricity bills by shifting their electricity usage to off-peak hours. This is an example of demand side management (DSM), a scheme where the electricity usage of consumers is changed by UCs by using various incentives.

### 2.2. Nonintrusive Appliance Load Monitoring (NALM)

NALM is a process for inferring which appliances have been used from the aggregated electricity load profile. Originally invented by Hart et al. [9] in 1985, today NALM is a family of methods all of which use different properties of the load signature in order to distinguish between electric appliances accurately.

NALM uses unique power signatures of appliances such as the (real and reactive) power levels, operation duration and periodicity to identify them. Appliances can in general be divided into three categories; namely ON/OFF appliances (such as a light bulb), Finite State Machine (FSM) appliances (such as refrigerators and washing machines) and continuously variable appliances (such as light dimmers and sewing machines). Unique properties of these appliance categories are utilized for an effective identification.

### 2.3. Notions of Privacy

Privacy threats that are introduced with the SG can be interpreted in different ways. As described earlier, NALM techniques can reveal the appliance usage data. However, some of this data can be of less importance in terms of privacy. For example; usage data of a toaster machine and a kettle can be interpreted as the resident being hungry, but how private this information is to the resident? These scenarios lead to different notions of privacy. NIST guidelines [11] listed some of the information which can be sensitive to the consumers as in Table 1. Some of these can be interpreted through the usage of metered load data.

*Table 1. Sensitive information made available by SGs [11].*

| Information | Description |
| --- | --- |
| Name | Name responsible for the account. |
| Address | Location to which service is being taken. |
| Meter Reading | kW, kWh power consumption collected at 15-60 minute intervals during the current billing cycle. |
| HAN | In-home electrical appliances. |
| Lifestyle | When the home is occupied and it is unoccupied, when occupants are awake and when they are asleep, how many various appliances are used, etc. |
| DER | The presence of on-site generation and/or storage devices, operational status, net supply to or consumption from the grid, usage patterns. |

To protect privacy, one approach can be to hide every electric signature regardless of its source. However, this approach can be inefficient as some of these signatures can be of less importance. Some researchers propose to hide only high frequency events [12]. The claim is that low frequency events such as refrigerators reveal less information about the private life of the consumer, while spikes in the load signatures can be of more personal appliances. Another privacy measure tries to prevent occupancy detection. The logic behind it is that the occupancy of consumer is an important information in terms of privacy, since it can be used for burglary by adversaries. For similar reasons, the information about the number of occupants and their genders can also be of importance.

# 3. PRIVACY PROTECTION METHODS

## 3.1. Noise Addition and Differential Privacy

Adding noise to the metered data is one of the privacy measures which is studied widely in the literature [13]-[20]. Most of these approaches try to achieve or maximize the *differential privacy* (DP), which is a statistical representation of the difference between the original and distorted data.

The main disadvantage of the noise addition is that it causes a mismatch between the metered load and the actual load. This prevents DSOs from rapidly and accurately reacting to outages, and UCs from producing accurate billing services as well as scheduling energy production in a precise manner. Therefore, noise-added consumption data undermines the very benefits of the SG. Access to the real metered data is crucial for setting accurate electricity prices as well as scheduling the energy production for better efficiency.

## 3.2. Data Aggregation and Cryptography

Data aggregation is another major method which aims to send aggregated multi-house power signatures to the UC so that an individual house's power signature can be hidden. Methods of homomorphic encryption are widely utilized in these studies [21]-[29].

The main drawback of these methods is the computational complexity of cryptographic techniques. Also, there is a need for a Trusted Third Party (TTP) for aggregation and/or encryption of multiple house data. This shifts the privacy concerns towards TTPs. Furthermore, metering multi-house power signatures can harm certain house-specific DSM applications, where each house is provided with individual incentives for changing their electricity usage patterns.

## 3.3. Data Anonymization and Downsampling

Data anonymization methods mainly focus on using pseudonyms instead of the customer IDs in order to protect their privacy [29], [30]. However, this approach does not protect customers from an adversary monitoring the electricity just outside of their houses.

Downsampling can simply be defined as reducing the frequency of metering energy data, so that small-duration appliance usage is hidden under the time-aggregated metered data. This method is shown to reduce the effectiveness of NALM methods [31] However; downsampling also hides the near-real-time electricity usage from the UCs and DSOs, reducing the effectiveness of the DSM applications and causing late responses to the failures.

## 3.4. Load Shaping

Load Shaping (LS) methods differ from the other approaches by sending the actual electricity usage data frequently and per-house basis. Therefore, LS eliminate the shortcomings of other methods. LS methods use household amenities (such as batteries, PEVs, RESs, shiftable appliances) in order to mask the real energy consumption of consumers. In the next section, the existing LS techniques are presented.

# 4. LOAD SHAPING BASED PRIVACY PROTECTION TECHNIQUES

## 4.1. Best Effort (BE)

Initially proposed by Kalogridis et al. [32], the Best Effort (BE) method tries to keep the load demand of the house as steady as possible with the help of a rechargeable battery (RB). This method results in a completely flat metered load if the capacity of the RB is big enough.

Three different privacy metrics are proposed in [32]: Relative Entropy (or Kullback-Leibler Divergence), clustering analysis and regression analysis. The efficiency of the method is evaluated by using these three metrics.

## 4.2. Non-intrusive Load Leveling (NILL)

McLaughlin et al. [33] proposed a LS method which uses a household RB to mask the appliance load signatures. Unlike BE method, where the metered load exposes appliance signatures when the battery is empty or full, NILL aims to maintain a steady load no matter the amount of energy stored in the RB. To achieve this, NILL adopts three target load levels. The algorithm chooses one of these levels according to the charging state of the battery. If the battery is close to full capacity and the load is below the steady state load $K_{ss}$, NILL chooses the *high recovery* state $K_H$ which sets the target metered load to a lower value. Therefore, the appliance load signatures are still masked, while the battery is not overcharged. Similarly, in *low recovery* state, the target metered load is set to a higher value so that the battery is charged while a constant metered load is maintained. The algorithm updates its $K_{ss}$ each time a low recovery or high recovery state is visited.

In [33], privacy performance is measured by the number of features (i.e. changes in the energy consumption between adjacent time slots) and empirical entropy.

### *4.3. Tolerable Deviation (TD)*

Tolerable Deviation (TD) strategy aims to keep the metered load *within* a certain limit of a target load. Introduced in [34], the idea is that small amounts of change in the metered load do not leak information about the appliances, and hence do not violate privacy.

In the study, the authors propose two different metrics for privacy: Number of (non-tolerable) changes in the metered load, and the mutual information between the actual load and the metered load [34].

### *4.4. Stepping*

Stepping method in [35] also utilizes a RB for LS. Similar to BE and NILL, this technique maintains a flat load until it is forced to change it. The change in the metered load is only allowed as a multitude of a certain step size.

The privacy performance of the stepping algorithms in [35] is measured in terms of the mutual information metric.

## 5. AMENITIES USED FOR LOAD SHAPING

### *5.1. Renewable Energy Sources (RESs)*

#### *5.1.1. Photovoltaic Energy (PVE)*

Photovoltaic energy (PVE) is one of the major sources of clean electricity. With the expansion of smart grid, more and more distributed PVE plants are expected to be integrated to the grid. With the decreasing cost of photovoltaic panels, the number of PVE producing houses is expected to increase [36].

In addition to reducing the electricity bill, PVE can also be used to mask some of the household energy usage. One shortcoming is, however, is the unpredictability of the PVE production. If the house is not equipped with a rechargeable battery (RB) or the capacity of the RB is not large enough, some of the excess PVE can be wasted.

#### *5.1.2. Wind Energy (WE)*

Another common RES which can be used for LS is wind energy. Wind energy is produced from the natural air flow by using a turbine to convert mechanical energy into electrical energy. It accounts for more than one percent of the overall electricity production worldwide [36]. SG enables distributed WE production by consumers. Therefore, WE can be used by households to mask the changes in their load demand. However, similar to PVE, WE is also unpredictable and in the absence of a RB it can only be used to mask the load at the instant it is generated.

### *5.2. Rechargeable Batteries (RB)*

Household batteries are expected to be a fundamental part of the Smart Homes (SHs) of the SGs. They can be used for privacy, as well as they are used for storing excess energy from RESs and for storing energy when it is cheaper in order to reduce the bill of the consumers.

However, using RBs impose a certain cost on the consumers [12]. They require a long time for an acceptable return of investment [37]. Also, the depreciation period of RBs are around 20 years [37], therefore it is not a one-time investment. Considering the installment and maintenance costs of RBs as well as their charging/discharging inefficiency are important when consumers invest in RBs either for profit or for privacy. Also, one should note that the capacity of RB has an impact of the effectiveness of LS methods. In order for LS methods to work efficiently, a large capacity RB is generally needed.

### *5.3. Shiftable Appliances*

Appliances in a household can be categorized into four types: time-shiftable, power-shiftable, time and power shiftable, and non-shiftable [38]. The operation time slots of time-shiftable appliances can be changed for load shaping in order to improve privacy at the cost of reducing the comfort of consumers. Similarly, power-shiftable appliances' power consumption can be modified to achieve the desired privacy. Note that some appliances such as Plug-in Electric Vehicles (PEVs) can be both time-shiftable and power-shiftable.

## 5.4. Plug-in Electric Vehicles (PEVs)

PEVs differ from the other (time and power shiftable) appliances in a house because they can draw energy from the grid *and* provide electricity for the other appliances thanks to their large-capacity internal batteries. Therefore, they can be used to shape the metered load for privacy just like a RB [39]. One important downside of using PEVs for privacy is that their primary purpose is different from that of RBs. A house can drain its RB for the purpose of cheap energy or privacy; however it may not be able to do so with PEVs because PEVs need to be sufficiently charged when the consumers need to use them.

## 5.5. Thermal Energy Storages - Combined Heat and Power

Majority of households are equipped with a form of thermal energy storage (TES), which is used either for heating or for supplying hot water to the house [40]. Therefore, TESs that are driven by electricity can be used as a thermal battery to hide load signatures. TESs are generally well-insulated, and hence energy efficient. However, a disadvantage of using TESs is that they convert energy in one direction only (electric energy to heat energy). Therefore, they can only be used to increase the load for certain time slots.

To mitigate the shortcomings of TESs, combined heating and power (CHP) devices are proposed in the literature [40]. The main advantage of CHPs is that they can generate both power and heat from various sources such as RESs and natural gas. In [40], cooling capability is also added to the CHP and the system is called combined cooling, heating and power unit (CCHP).

## 6. PRIVACY PERFORMANCE METRICS

### 6.1. Information Theoretic Metrics

#### 6.1.1. Mutual Information (MI)

In information theory, the mutual information (MI) is a metric between two sets of data which quantifies how much information of a set of data can be obtained by observing the other. In case of smart grid privacy, the mentioned two sets of data are the metered load and the actual load. Many studies including [34], [35] and [41] utilize the MI as a performance metric. However, the rationale behind using MI as a metric is not clear, as NALM methods use various techniques such as edge detection and clustering to identify individual appliances, and having a lower MI does not always guarantee that the load signatures of appliances are hidden.

#### 6.1.2. Relative Entropy (RE)

Also known as Kullback-Leibler Divergence (KLD), relative entropy (RE) is also a metric which is used to distinguish a probability distribution from another. It measures the expected number of bits (of data) lost between a distribution and another distribution, which is the approximation of the first one. In case of measuring privacy, the first distribution is the actual load and the second distribution is the metered load. Thus, KLD gives us the information lost by changing the actual load with the metered load. Many studies including [32] and [42] use RE as a performance metric.

### 6.2. Statistical Metrics and Correlation Coefficients

#### 6.2.1. Mean Square Error (MSE)

In statistics, mean square error (MSE) measures the average of the squared distances between a set of data and its estimation. In the smart grid literature, it is used as a privacy metric where the set of data is the actual consumption and the estimation is the metered load [43]. Although it measures the distance between the metered and the actual load, a high MSE does not guarantee a high privacy, as privacy is not directly related with the distance between the two sets of data.

#### 6.2.2. Coefficient of Determination (CoD)

Various correlation coefficients (CCs) such as Fisher correlation coefficient (FCC), Matthews correlation coefficient (MCC) and Pearson correlation coefficient (PCC) are also used for measuring privacy in smart grids [32], [44]. Similar with the CoD, CCs also measure the relationship between two sets of data. The difference between CCs and the CoD is that correlation coefficients measure the linear correlation between two variables. In case of smart grid privacy, they are used to measure the correlation between the metered load and the actual load.

### *6.3. Empirical Metrics*

#### *6.3.1. Number of Changes (NoC)*

Since NALM methods generally look for the changes in the load signature and match them with the change in the load caused by individual appliances, some of the studies use the number of changes (NoC) in the load signature between each time slot in order to measure the performance of their methods [10], [33]. In this case, 0 is the best possible outcome, which indicates a completely flat load. The more changes there are in the load, the worse the method's performance in terms of privacy is. However, this metric is not able to measure privacy directly, similarly to the previously mentioned metrics.

#### *6.3.2. Cluster analysis*

Cluster analysis is the grouping a set of data such that the data in the same group are more similar to each other, compared to the data from the other groups. In [32], cluster analysis is used to measure the privacy, where clustering is made by using the changes in the actual load data first and then by using the changes in the metered load; and the number of correct classifications in the latter one is counted. The more correct classifications there are, the worse the performance is. This metric is one of the most accurate ones for measuring privacy, as it runs a NALM method on the actual and metered load and calculates the difference. However, this metric is computationally more complex than the other metrics, as it employs machine learning techniques for clustering. In addition, there are numerous NALM methods in the literature, hence having a high privacy against cluster analysis cannot indicate high privacy against all of the NALM methods.

### *6.4. Combined Metrics*

In [7], a combination of three performance metrics, namely NoC, CoD and RE, is used to derive a combined performance metric. This approach can be beneficial for an accurate measurement of privacy, as none of the metrics alone can guarantee an accurate measurement. However, further analysis is needed to assess if using the combination of more than one metrics yields in a more accurate measurement of privacy than those obtained by using one of the singular metrics.

## 7. POSSIBLE RESEARCH DIRECTIONS FOR PRIVACY IN SMART GRIDS

### *7.1. Definition of Privacy*

As discussed in this paper, there are various information that can be inferred from monitoring the load demand of the consumers. Some of this information can be sensitive to a type of consumer, whereas another consumer may not be interested in protecting them. Therefore, which information is important to which type of customer should be assessed and the research should focus on protecting this particular information. This is where sociology and power systems engineering disciplines can join forces. An example study in the field of sociology is presented in [47], where authors try to assess which type of information that people want to be protected against adversaries in case of SGs.

### *7.2. Commercial Privacy vs. Residential Privacy*

Another issue that should be addressed is the privacy needs of commercial or industrial consumers and residential customers. An industrial or commercial company might want to hide different information compared to the residential users (such as shift hours, product design changes etc.). Also, their adversaries can be very different than that of residential consumers (such as rival companies, intelligence agencies, etc.). Therefore, privacy for commercial and industrial customers should be considered separately from the residential ones.

### *7.3. One Privacy Measure for All*

The literature on smart grid privacy does not use a single privacy measure as discussed earlier. This is partly because none of the existing metrics is able to accurately measure protection against NALM techniques. Coming up with an accurate privacy measure for smart grid privacy methods is an open research area.

*7.4. Opting out of Smart Grid*

The acceptance of the smart grid by the public is an important issue, as this new technology arises new questions in peoples' minds, especially in terms of privacy [48]. Some EU countries such as Netherlands integrate an option for the consumers to "opt-out" from smart grid, which means they can turn on and off smart grid functionality as they want [49]. This flexibility comes with complexity in both the design and the operation of the smart grid. Current literature assumes that all of the consumers are part of the smart grid scheme. Therefore, the issues that arise with dynamic customers are yet to be addressed by the research community.

## 8. CONCLUSION

Smart grid comes with many advantages, thanks to its ability to convey both power and information between its constituents. Detailed power usage data of consumers enable a more efficient power production as well as customized and more profitable services for the consumers. On the other hand, high frequency metered load data can be used by adversaries to break it down to individual appliance usage data, which can be used against the consumers' consent. Therefore, there are many techniques proposed in the literature to mitigate this privacy problem.

This paper presented an overview of existing privacy protection techniques, as well as the means that can be used for ensuring privacy and the privacy metrics. A categorization of existing techniques was provided along with their advantages and disadvantages. The privacy metrics used in the literature and the logic behind using them were explained. Household amenities that can be used for load shaping and hence providing privacy were presented. Possible research directions concerning privacy protection in smart grids were also provided in order to shed light upon the future of the topic.